\documentclass{article}
\usepackage{graphics}
\usepackage{latexsym}
\newcommand{\bw}{\mathbf{w}}
\newcommand{\R}{\mathbf{R}}
\newcommand{\be}{\begin{equation}}
\newcommand{\ee}{\end{equation}}
\newcommand{\Flim}{\mathop{F\mbox{-}\lim}}
\newcommand{\floor}[1]{\lfloor #1\rfloor}
\newtheorem{definition}{Definition}
\newtheorem{theorem}[definition]{Theorem}
\newtheorem{lemma}[definition]{Lemma}

\title{Calculus on Fractal Curves in~$\mathbf{R}^n$}
\author{Abhay Parvate$^{1,2}$ \and Seema Satin$^1$ \and A.D.Gangal$^1$\\
\and $^1$Department of Physics, University of Pune, Pune 411~007, India \\
\and $^2$Center for Modelling and Simulation, University of Pune, Pune 411~007, 
India \\
\and \texttt{abhay@physics.unipune.ernet.in }, 
\and \texttt{satin@physics.unipune.ernet.in},
\and \texttt{adg@physics.unipune.ernet.in}
}
\date{ }
\begin{document}
\maketitle
\begin{abstract}
A new calculus  on fractal curves, such as the von Koch curve, is formulated. 
We define a Riemann-like integral along a
fractal curve $F$, called $F^\alpha$-integral, where $\alpha$ is
the dimension of $F$.
 A derivative along the fractal curve called $F^\alpha$-derivative, is also 
defined.
The mass function, a measure-like algorithmic quantity on the curves, plays a 
central role in the formulation. An appropriate algorithm to calculate the mass
function is presented to emphasize its algorithmic aspect.

Several aspects of this calculus retain much of 
the simplicity of ordinary calculus. 
We establish a conjugacy between this
calculus and ordinary calculus on the real line. The $F^\alpha$-integral and 
$F^\alpha$-
derivative are shown to be conjugate to the Riemann integral and ordinary 
derivative respectively. In 
fact, they can thus be evalutated using
the corresponding operators in ordinary calculus and conjugacy.
Sobolev Spaces are constructed on $F$, and 
$F^\alpha$- differentiability is generalized .
Finally we touch upon an  example of absorption along fractal paths, to
illustrate the utility of the framework in model making.
\end{abstract}
\section{Introduction}
\label{sec-introduction}
It is now well known that fractals pervade nature \cite{Mandelbrot,Bunde}. The
 geometry of fractals
is also well studied \cite{Mandelbrot,Falconer,Falconer1,Falconer2,Tricot,
Edgar}. Fractal curves often lack the smoothness properties required by ordinary calculus. 
For 
example, observed path of a quantum mechanical particle \cite{Abbott} or 
Brownian  and Fractional Brownian
trajectories \cite{Mandelbrot,Falconer} are known to be fractals and are 
continuous but 
non-differentiable. A percolating path, just above the percolating phase
 transition
can be considered as an appoximate realization of a fractal curve 
\cite{Havlin}. If a long polymer is modeled as a fractal curve, then 
accumulation of a physical property along the curve would amount to 
integration on such a curve. This is often carried out using ad hoc procedures.

While there are some remarkable approaches to develop tools for such situations \cite{Adrover, Schwalm, Kigami, Giona, Mendivil}, much more is desired.
This paper aims to formulate a calculus specifically tailored for 
fractal curves, in a close analogy with ordinary calculus.
In particular, we adopt a Riemann-Stieltjes like approach for defining integrals, 
because of 
its simplicity and advantage from algorithmic point of view. Such an approach 
was concieved in 
 \cite{Kolwankar} which began with formulation in the terms of Local Fractional
 derivatives. A new prescription was proposed to give meaning to differential 
equations on Cantor-like sets which are totally disconnected where the Local
Fractional Derivatives do not carry over. The calculus formulated and developed
in \cite{Abhay,Abhay00,Abhay01} for fractal subsets of $\mathbf{R}$ fully 
justifies the prescription in \cite{Kolwankar}.
 In particular, an integral and a derivative of order $\alpha$
are defined \cite{Abhay} on Cantor-like totally disconnected subsets of the 
real line, where $\alpha \in (0,1]$
is the dimension of $F$. This calculus, called $F^\alpha$- calculus has many
results analogous to ordinary calculus and can be viewed as a generalization
of ordinary calculus on $\R$. In fact,
in \cite{Abhay00,Abhay01} a conjugacy between the $F^\alpha$-calculus and
 ordinary calculus is discussed. 

The present paper extends that approach, which was developed for disconnected 
sets like Cantor-sets, to formulate calculus on fractal curves which are
continuous
. The organization of the paper is as follows.
In Section~\ref{sec-staircase} we define a mass function and integral staircase
 function. The mass function
gives the content of a continuous piece of the fractal curve $F$. The staircase 
function, more appropriately called the "rise function", is obtained from the
 mass
function and describes the rise of the mass of the curve with respect to 
the parameter. 
We emphasize the algorithmic nature of the mass function: by presenting an 
algorithm to calculate it. 
In section \ref{sec: gammadim} we show that the mass
function allows us to define a new dimension called $\gamma\mbox{-}dimension$,
 which is algorithmic and finer than  the box dimension.
 In section \ref{sec:algorithm} we discuss the 
algorithmic nature of mass function and present an algorithm to calculate it. 
In section \ref{sec-f_continuity}
the concepts of limits and continuity are adapted to the concepts of $F$-limit
 and $F$-continuity. Section \ref{sec-falpha-int} is devoted to the discussion
 of integral
on fractal curves called $F^\alpha$-integral. The formulation is analogous to
the Riemann integration \cite{Goldberg}.
The notion of $F^\alpha$-differentiation is introduced 
in section \ref{sec-alphadiff}.
The fundamental theorems of $F^\alpha$-calculus proved in section
 \ref{sec-fundamental}, state that the $F^\alpha$-integral and $F^\alpha$-
derivative are inverses of each other.  
The conjugacy between $F^\alpha$-calculus  on $F$ and ordinary calculus on the
real line,
discussed in section \ref{sec:conjugacy}, establishes a relation between the two and
gives a simple method to evaluate $F^\alpha$-integrals and $F^\alpha$-
derivatives of functions on the fractal $F$.
In section \ref{sec:funcspace},  
function spaces of $F^\alpha$-integrable and $F^\alpha$-differentiable
functions on the fractal $F$ are explored. In particular Sobolev Spaces are
introduced and abstract Sobolev derivatives are constructed. Finally as a 
simple physical application we 
briefly touch upon, as an example, a simple model of absorption along a fractal
path in section \ref{sec:absorption}. Section \ref{sec:conclusion} is the
 concluding section.  
\section{The mass function and the staircase}
\label{sec-staircase}

This paper can be considered as a logical extension of calculus on fractal 
subsets of the real line developed in \cite{Abhay01}.
The proofs which are analogous to those in \cite{Abhay01} are omitted. 

In this paper we consider fractal curves, i.e. images of continuous functions
$f: \R \rightarrow \mathbf{R^n}$ which are fractals. To be precise:

  Let $[a_0,b_0]$
be a closed interval of the real line.  

\begin{definition} A fractal (curve) $F \subset \mathbf{R^n}$ is said to be 
continuously 
parametrizable (or just parametrizable for brevity) if there exists a function
$\mathbf{w}:[a_0,b_0] \rightarrow F \subset \mathbf{R^n}$ which is continuous, 
one-to-one and onto  
$F$.
\end{definition}
 In this paper $F$ will always denote such a fractal 
curve.

\textbf{Examples:}
\begin{enumerate}
\item A simple example of such a parametrization is the function $\bw :\R 
\rightarrow \R^2$ defined by $\bw(t) = (t, W^s_\lambda (t))$ where $W^s_\lambda
(t)$
is the well known Weierstrass function \cite{Falconer} given by
\[W^s_\lambda (t) = \sum_{k=1}^\infty  \lambda^{(s-2)k} \sin \lambda^k t\]
where $\lambda > 1$ and $1<s<2$. The graph of $W^s_\lambda(t)$ is 
known to be a fractal curve with box-dimension $s$.

\item Our next example constitutes of one important class of parametrizations
 of self-similar curves in two
dimensions (There are other ways of parametrizing fractal curves
; for example see \cite{Schnieder}). Let $T_i, i = 0, \dots, n-1$ be linear 
operations 
which are
composed of rotation and scaling. Each $T_i$ can be represented by a $2
\times 2$ matrix:
\[
        T_i = s_i
        \left[
        \begin{array}{rr}
                \cos \theta_i & - \sin \theta_i \\
                \sin \theta_i &  \cos \theta_i
        \end{array}
        \right].
\]

Further, they should satisfy the condition:
\[
        \sum_{i=0}^{n-1} T_i(\mathbf{v}) = \mathbf{v}
\]
for any vector $\mathbf{v}$, and $0 < s_i < 1$ for $i = 0,\dots,n-1$.
The fractal is defined by the limit set \cite{Falconer1} of the similarity
transformations:
\[
        S_j(\mathbf{v}) = \sum_{i=0}^{j-1} T_i(\mathbf{v_0}) +
        T_j(\mathbf{v}), \qquad
        j = 0, \dots, n-1
\]
where $\mathbf{v_0}$ is a fixed vector.The limit set will be in the form of a
curve because of the way $S_j$ are constructed from
$T_i$.

Let $\floor{nt}$ denote the integer part of $ nt $. Now, the function 
$\mathbf{w}$ defined implicitly by 
\be \label{eq:transformation}
        \bw(t) =
        \sum_{i=0}^{\floor{nt} - 1} T_i(\mathbf{v_0}) +
        T_{\floor{nt}}(\bw(nt - \floor{nt})), \qquad
        0 \leq t \leq 1
\ee
parametrizes the above fractal curve. 
 To implement it as an algorithm, we stop the recursion
at some appropriate depth. 
The continuity and invertibility of this parametrization can be numerically
verified, when the curve itself is non-self-intersecting.

In particular the von Koch curve is realized by setting all $s_i = 1/3$,
$\theta_0 = \theta_3 = 0$, $\theta_1 = -\theta_2 = \pi/3$, and $\mathbf{v_0}
= (1,0)$ (the unit vector along $x$~axis).

$\bullet$
\end{enumerate} 
Hereafter symbols such as $a$,$b$,$c$,etc denote numbers
in $[a_0,b_0]$ and $\theta$, $\theta'$ etc denote points of $F$. 
\begin{definition} For a set F and a subdivision $P_{[a,b]}, a<b$, $[a,b]
\subset [a_0,b_0]$
\be \sigma^{\alpha}[F,P] = \sum_{i=0}^{n-1} \frac{|\mathbf{w}(t_{i+1}) - 
\mathbf{w}(t_i)|^\alpha}
{\Gamma(\alpha +1)}  \label{eq:sigma}\ee
where $|\cdot|$ denotes the euclidean norm on $\mathbf{R^n}$, $1 \leq \alpha
\leq n$ and $P_{[a,b]} = 
\{a=t_0,\ldots,t_n=b\}$. 
\end{definition}
Next we define the coarsed grained mass function.
\begin{definition}\label{def:gamma-delta} 
Given $\delta >0$ and $a_0 \leq a\leq b \leq b_0$, the coarse grained mass 
$\gamma_\delta^\alpha(F,a,b)$ is given by
\be \gamma_{\delta}^{\alpha}(F,a,b) = \inf_{\{P_{[a,b]}:|P| \leq \delta\}}
\sigma^{\alpha}[F,P] \label{eq:gamma}\ee 
\[ \mbox{ where }|P| = \max_{0\leq i \leq {n-1}}(t_{i+1}-t_i)
\mbox{ for a subdivision } P. \]  
\end{definition}

The mass function is the limit of the coarse-grained mass as $\delta 
\rightarrow 0 $ :
\begin{definition} For $a_0 \leq a \leq b \leq b_0 $, the mass function 
$\gamma^\alpha(F,a,b)$ is given
by
\[ \gamma^\alpha(F,a,b)= \lim_{\delta \rightarrow
0}\gamma_\delta^\alpha(F,a,b) \]

Remark: Since $\gamma$ is a monotonic function of $\delta$.
 The limit exists , but could be finite or $+ \infty$.
\end{definition}
The following properties of the mass function follow easily.
\subsection*{Properties of $\gamma^\alpha(F,a,b)$}
\begin{itemize}
\item For $a_0 \leq a < b < c \leq b_0 $ and $ \gamma^{\alpha}(F,a,c) < \infty
 $

\[ \gamma^{\alpha}(F,a,c)= \gamma^{\alpha}(F,a,b) + \gamma^{\alpha}
(F,b,c). \label{eq:addgamma}\]

\item $\gamma^\alpha(F,a,b)$ is increasing in $b$ and decreasing in
 $a$.
\item If $\gamma^\alpha(F,a,b)$ is finite, $\gamma^\alpha(F,a,t)$ is continuous
for $t \in [a,b]$.

\textit{Remark}: The implication of this result is that no single point has a 
nonzero mass, or in other words, the mass function is atomless.  

\item Let $F \subset \R^n$ be  parametrizable.  Let $\lambda$
be a positive real number, $\mathbf{v} \in \R^n$, and let $T$ be a rotation 
operator. We denote

\[ F + \mathbf{v} =\{ \mathbf{w}(t) + \mathbf{v} : t \in [a_0,b_0] \} \]

\[\lambda F = \{\lambda \mathbf{w}(t):t \in [a_0,b_0] \}. \]
and \[T F = \{T \mathbf{w(t)} : t \in [a_0,b_0]\} \]
Then,
\begin{enumerate}
\item Translation :
\[\gamma^\alpha(F + \mathbf{v},a,b) = \gamma^\alpha(F,a,b)\]
\item Scaling :
\[\gamma^\alpha(\lambda F,a,b) = \lambda^\alpha \gamma^\alpha(F,a,b)\]
\item Rotation :
\[\gamma^\alpha(TF,a,b)= \gamma^\alpha(F,a,b)\]
\end{enumerate}
\end{itemize} 
\subsection*{Re-parametrization Invariance of Mass Function}

The definitions of $\sigma^\alpha$, $\gamma^\alpha_\delta$, and therefore
$\gamma^\alpha$ implicitly involve the particular
parametrization~$\bw$. Here we show that although defined through the
parametrization, these definitions are invariant under the change of
parametrization. In order to be able to unambiguously and explicitly refer
to the parametrization, we introduce a temporary change in the notation to 
explicitly indicate dependence on parametrization.
Thus given a parametrization~$\bw : [a, b] \rightarrow \R^n$, we use the
following notation here:
\begin{eqnarray*}
    \sigma^\alpha[F, P; \bw]
        &=&
        \sum_{i=0}^{n-1}
        \frac{|\bw(t_{i+1}) - \bw(t_i)|^\alpha}{\Gamma(\alpha+1)}\\
    \gamma^\alpha_\delta(F, a, b; \bw)
        &=&
        \inf_{|P| \leq \delta} \sigma^\alpha[F, P; \bw]\\
    \gamma^\alpha(F, a, b; \bw)
        &=&
        \lim_{\delta \rightarrow 0}
        \gamma^\alpha_\delta(F, a, b; \bw)
\end{eqnarray*}

Let $\bw_1$ and $\bw_2$ be two parametrizations of the given fractal curve.
By our definition of parametrization, $\bw_1$ and $\bw_2$ are continuous and
one-to-one.
Let the domain of $\bw_1$ be $[a_1, b_1]$, and that of $\bw_2$
be $[a_2, b_2]$.
We further assume that $\bw_1$ and $\bw_2$ have the same
\emph{orientation}, i.~e.\
$\bw_1(a_1) = \bw_2(a_2)$
and
$\bw_1(b_1) = \bw_2(b_2)$.
Thus, $z = \bw_2^{-1} \circ \bw_1 : [a_1, b_1] \rightarrow [a_2, b_2]$ is a
continuous, one-to-one and strictly monotonically increasing function.

Now, given $\delta_2 > 0$ and $\epsilon > 0$, there exists a
subdivision~$P_2$ of $[a_2, b_2]$ such that
\[
    \sigma^\alpha[F, P_2; \bw_2] <
    \gamma^\alpha_{\delta_2}(F, a_2, b_2; \bw_2) + \epsilon.
\]
The set of points $P_1 = \{z^{-1}(t): t \in P_2\}$ forms a subdivision of
$[a_1, b_1]$.  Then,
\[
    \sigma^\alpha[F, P_1; \bw_1] = \sigma^\alpha[F, P_2; \bw_2]
\]
by appropriate substitution. Therefore,
\[
    \sigma^\alpha[F, P_1; \bw_1] <
    \gamma^\alpha_{\delta_2}(F, a_2, b_2; \bw_2) + \epsilon
\]
which implies that
\[
    \gamma^\alpha_{\delta_1}(F, a_1, b_1; \bw_1) <
    \gamma^\alpha_{\delta_2}(F, a_2, b_2; \bw_2) + \epsilon
\]
where $\delta_1 = |P_1|$.
Further, since $z$ is continuous,
$\lim \delta_1 = 0,\mbox{ as } \delta_2 \rightarrow 0$,
implying that
\[
    \gamma^\alpha(F, a_1, b_1; \bw_1) <
    \gamma^\alpha(F, a_2, b_2; \bw_2) + \epsilon.
\]
Since $\epsilon$ is arbitrary, and the same argument remains valid starting
with $z^{-1} = \bw_1^{-1} \circ \bw_2$, we conclude that
\[
    \gamma^\alpha(F, a_1, b_1; \bw_1) =
    \gamma^\alpha(F, a_2, b_2; \bw_2).
\]
This establishes the fact that the mass function depends only on the
fractal curve (i.~e.\ the image of the parametrization), and is independent
of the parametrization itself.
Since the mass function underlies the calculus developed in the subsequent
sections, the calculus is also independent of the particular
parametrization chosen.

Now we introduce the integral staircase function for a set $F$ of order
$\alpha$. 
\begin{definition} Let $p_0 \in [a_0,b_0]$ be arbitrary but fixed. The 
staircase function $S_F^\alpha: [a_0,b_0]\rightarrow
\mathbf{R}$ of order $\alpha$ for a set F is given by
\be
S_F^{\alpha}(t) = \left\{ \begin{array}{ll}
	\gamma^{\alpha}(F,p_0,t) & t \geq p_0 \\
	- \gamma^\alpha(F,t,p_0) & t < p_0
		\end{array}
	\right. 
	\label{eq:staircase_function} \ee
where $t \in [a_0,b_0]$.
\end{definition}
In the rest of this paper we take $p_0=a_0$ unless stated otherwise.

Here this function may, more appropriately, be described as a rise function.
However we retain the name staircase function because in analogous calculus on
fractal subsets of the real line this role is played by a staircase.

Throughout the paper we consider only those sets for which $S_F^\alpha$ is 
strictly increasing and thus invertible. Further, we define 
\be
 J(\theta)= S_F^\alpha(\mathbf{w}^{-1}(\theta)), \quad \theta \in F
\ee
 which is the function induced by 
$S_F^\alpha$ on $F$, and it is also one-to-one.

\begin{figure}[htb]
\includegraphics{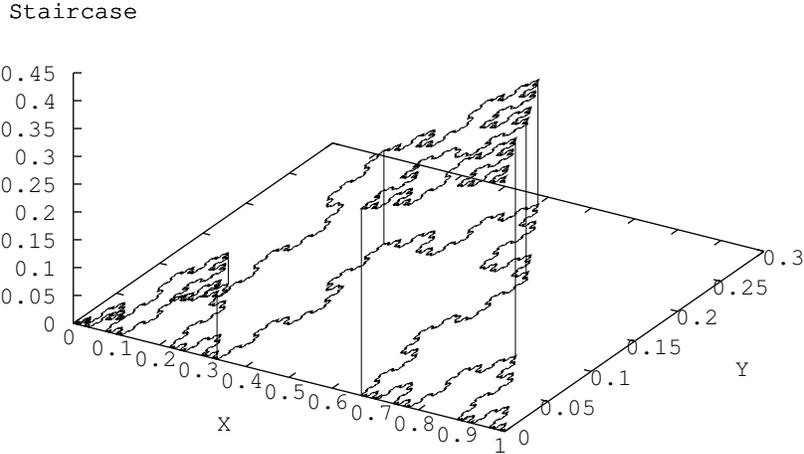}
\caption{$S_F^\alpha$ for von Koch curve. The von Koch curve lies in the $XY$ 
plane.
The vertical lines are drawn to guide the eye (to show how $S_F^\alpha$ rises)}
\label{fig:staircase}
\end{figure}

As an example, figure \ref{fig:staircase} shows the staircase function for the 
von koch curve. The curve was parametrized as given in \cite{Schnieder}.

\begin{figure}[htb]
\includegraphics{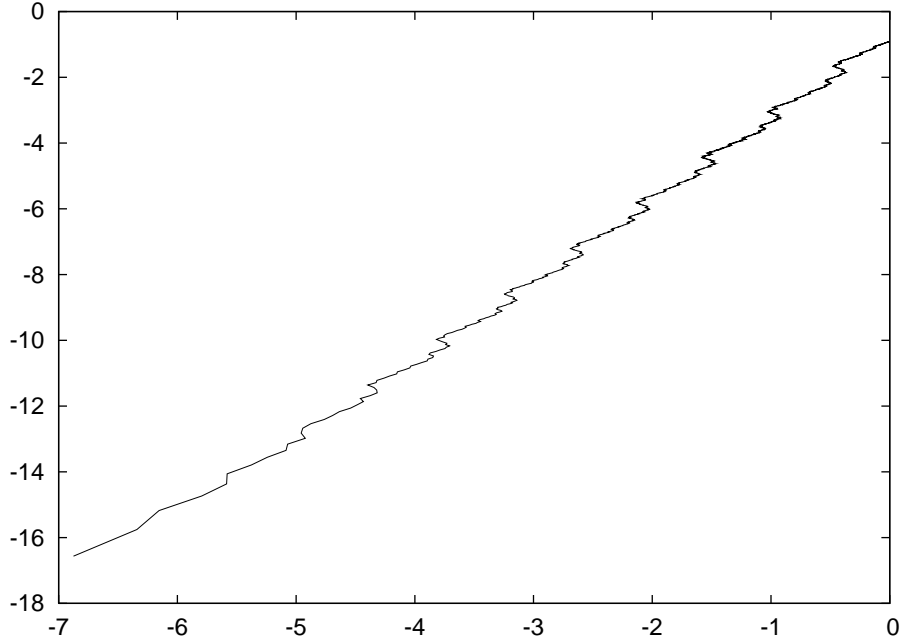}
\caption{log-log graph of Euclidean distance between origin and $\mathbf{w}(t)$
 (Y-axiz) vs $S_F^\alpha(t)$ for 
$t \in [0,1]$  for von-koch curve}
\label{fig:graph2}
\end{figure}

A  log-log graph of the staircase  function $S_F^\alpha(t)$ against the 
Euclidean distance between origin and $\mathbf{w}(t)$ for
the von- koch curve is shown in fig ~\ref{fig:graph2}.

\section{The $\gamma$- Dimension} \label{sec: gammadim}
We now consider the sets $F$ for which the mass function $\gamma^\alpha
(F,a,b)$ gives the most useful information. Due to the similarity of the 
definitions of mass function and the Hausdorff outer measure, the
 former can be used to define a fractal dimension as follows.

It can be seen that $\gamma^\alpha(F,a,b)$ is infinite upto certain value of 
$\alpha$, say $\alpha_0$, and jumps down to zero for $ \alpha > \alpha_0$. Thus 
\begin{definition}
\label{def:dimension}
The $\gamma$-dimension of $F$, denoted by $\dim_\gamma(F)$, is
\[\dim_\gamma(F) = \inf\{\alpha : \gamma^\alpha(F,a,b) = 0\} = \sup
\{\alpha : \gamma^\alpha(F,a,b) = \infty \}\]
\end{definition}
It follows that the $\gamma$-dimension is finer than the box dimension. Thus
\[\dim_\gamma(F)\leq \dim_B(F).\]
\subsection*{$\gamma$-dimension for self-similar curves}
Let $\alpha$ denote the $\gamma$- dimension of a self similar curve ,
which is made up of $m$ copies of itself, scaled by a factor of $\frac{1}{n}$
and rotated and translated appropriately. 
Then using the translation, scaling and rotation properties of the mass 
function, one can see that the mass of the whole
 curve is given by
\[
\gamma^\alpha(F,a_0,b_0) = m \gamma^\alpha(\frac{1}{n} F,a_0,b_0)
\]
\be
\gamma^\alpha(F,a_0,b_0) = m (\frac{1}{n})^\alpha \gamma^\alpha(F,a_0,b_0)
\ee

Hence, 
\be
\alpha = \log{m}/\log{n} 
\ee
This is same as the Hausdorff dimension of self-similar curves \cite{Falconer1}
.

 Thus for self-similar curves
\[\dim_\gamma F = \dim_{\mathcal{H}} F = \dim_B F\]
where $dim_{\mathcal{H}} F$ denotes the Hausdorff dimension and $dim_B F $ the 
box dimension of $F$.
\section{Algorithmic Nature of the Mass Function} \label{sec:algorithm}
Let us first summarize the definition of mass function:
\begin{equation}
    \gamma^\alpha(F, a, b) =
    \lim_{\delta \rightarrow 0}
    \inf_{\{P: |P| \leq \delta\}}
    \sum_{i=0}^{n-1}
    \frac{|\bw(t_{i+1}) - \bw(t_i)|^\alpha}{\Gamma(\alpha+1)}
    \label{eqn:algo-definition}
\end{equation}
One of the main difference between the Hausdorff measure and the mass
function is that while the Hausdorff measure is based on sums
over a countable covers (composed of arbitrary sets) of the given set $F$,
the mass function is based on finite subdivisions of the parametrization
domain. From an algorithmic point of view, the extent of the set of all
possible finite subdivisions is much smaller than that of all countable
(finite and infinite) covers of a set. This makes the mass function much
more amenable to an algorithmic computation.

As in any algorithm which intends to approximate the infimum, we would like 
 to find a subdivision $P$ such that
$\sigma^\alpha[F, P]$ is close to the infimum. Further, we can
consider values of $\delta$ only as small as practically possible within
the reach of numerical calculations. The goal of the algorithm 
is thus to find a subdivision $P$ as described above, given a fixed
$\delta$.

However, the set of allowed subdivisions is still large, to explore all of it
systematically. Further the constraint $|P| \leq \delta$ does not restrict
the number of points in $P$, rendering the standard deterministic
optimization algorithms either inapplicable or too complex to implement.
More appropriate is a Monte Carlo method where a subdivision is modified in
a variety of ways randomly but consistently with the constraint $|P| \leq
\delta$, and the change is accepted if the sum $\sigma^\alpha[F, P]$
decreases due to the modification.
The algorithm presented below, is based on this strategy.

\begin{figure}[htb]
\includegraphics{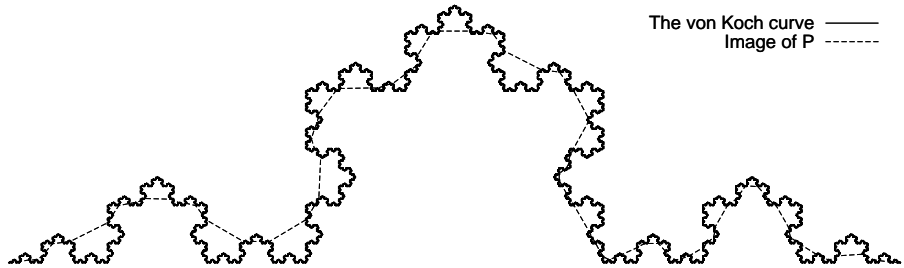}
\caption{The image (under $\bw$) of a numerically computed near-optimal
subdivision~$P$, for $\delta = 0.05$, superimposed on the von-Koch curve.
}
\label{fig:opt-subdivision}
\end{figure}

\begin{figure}[htb]
\includegraphics{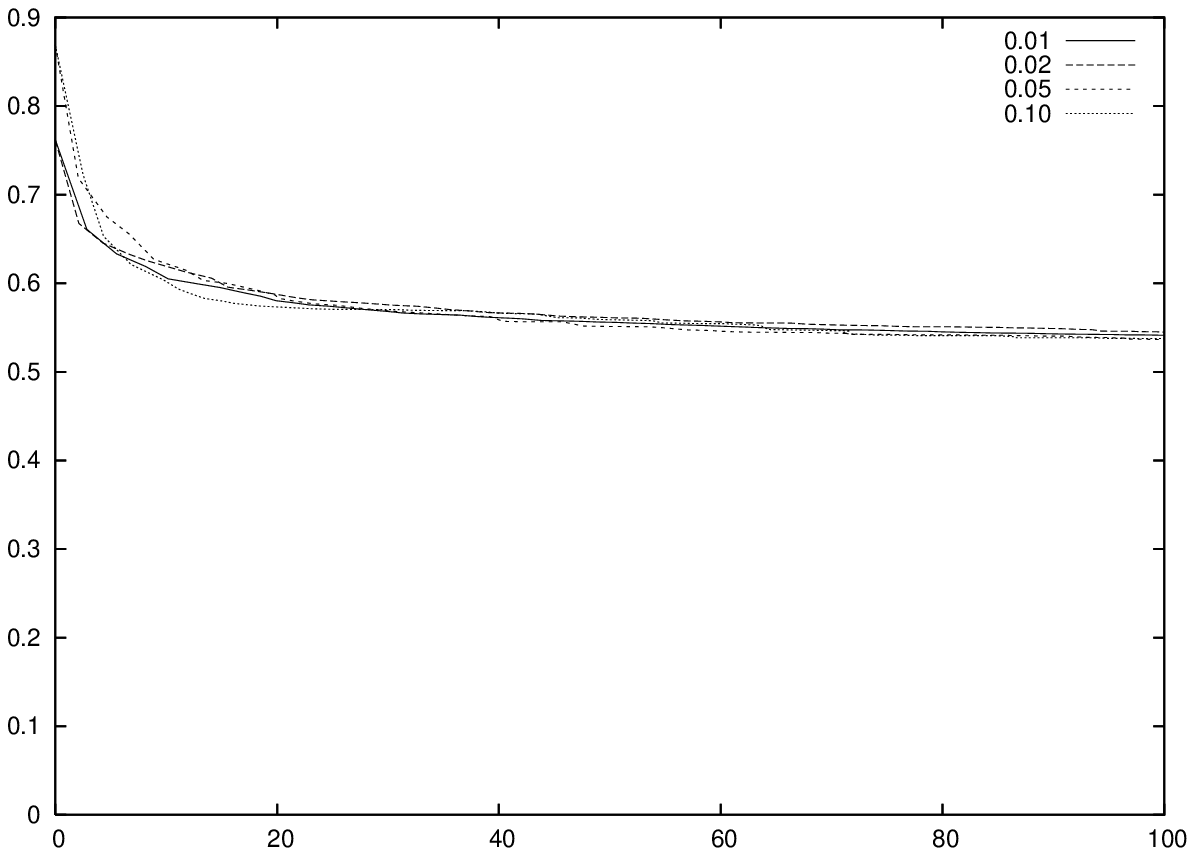}
\caption{
The evolution of $\Gamma(\alpha +1).\sigma^\alpha[F, P]$ over the normalized
 number of
iterations. The evolution is shown only up to $N' = 100$, since the latter
part ($100 < N' \leq 2000$) is almost flat and uninteresting.
}
\label{fig:evol}
\end{figure}

\subsection*{A Monte Carlo Algorithm}


For the purpose of this algorithm, $[a,b]$ denotes the domain of $\bw$.
Further, ``randomly'' means with a uniform probability unless stated otherwise.
The symbol $P$ always indicates the ``current'' subdivision in
consideration.

We begin with a \emph{uniform} subdivision $P$ such that $|P| = \delta/4$,
and iteratively improve it using the following prescription. 

\begin{enumerate}
\item
    Choose two numbers $x, y \in [a,b]$ randomly, and relabel them if
    necessary so that $x \leq y$. Then $[x, y] \subset [a,b]$.
    Let $P' = \{t_i: 0 \leq i \leq m\}$ denote the set of all points of
    $P \cap [x, y]$. We now modify $P'$ in one of the following ways with
    equal probability, and denote the resultant by $P''$:
    \begin{enumerate}
    \item
        With a probability $p_c = \min(1, \delta/(y-x))$,
        we shift each point $t_i$ (except $t_0$ and $t_m$)
        by a random amount between
        $[-\delta/2, \delta/2]$, if the resultant subdivision
        $P''$ still satisfies $|P''| \leq \delta$.
    \item
        With a probability $p_d = \min(1, \delta/(y-x))$,
        we remove each point $t_i$ (except $t_0$ and $t_m$)
        from $P'$, if the resultant subdivision
        $P''$ still satisfies $|P''| \leq \delta$.
    \item
        With a probability $p_i = \min(1, \delta/(y-x))$,
        we insert a point between each $t_i$ and $t_{i+1}$ which is chosen
        randomly from $[t_i, t_{i+1}]$. (However, to avoid accumulating
        too much of rounding error, we insert the point only if the
        distance between  $t_i$ and $t_{i+1}$ is greater than $\delta/10$.)
    \end{enumerate}
\item
    Form a new subdivision
    $P_1 = (P \cap [a, x)) \cup P'' \cup (P \cap (y, b])$, i.~e.\ the
    subdivision of which the points belonging to $[x, y]$ are changed by
    the above procedure.
    If $\sigma^\alpha[F, P_1] < \sigma^\alpha[F, P]$, then we consider
    $P_1$ as the ``current'' subdivision which will be possibly improved
    further using above steps. Otherwise we consider $P$ again for the
    purpose.
\end{enumerate}

As the sum $\sigma^\alpha[F, P]$ approaches the infimum, many of the
newly formed subdivisions $P'$ are rejected since they sum up higher than
$P$. Thus near the infimum, the sum remains constant for many consecutive
iterations, and changes only intermittently. Therefore the usual convergence
criterion of terminating iteration when the difference between successive
iterations or every $K$ iterations ($K$ being a suitable
large integer) goes below certain small number, is not useful in this case.
Instead,
after examining the sum over a large number of iterations,
we observe
that the sum stops making significant progress between $N' = 1000$ to
$N'=2000$, where $N' = N/n$
is the number of iterations $N$ \emph{normalized} by the current
subdivision size~$n$. Further, we need to go through all these iterations
more than once, just to ensure that subdivision is really optimal. Occasionally
it may happen that the sum settles a little above the optimal value, gettting
''trapped'' in a ''local minimum''.


We demonstrate the results of this algorithm as applied on the von Koch
curve, parametrized as in equation~(\ref{eq:transformation}). It turns out that
the mass of the entire von Koch curve is a little less than
$0.51/\Gamma(\alpha+1)$, $\alpha=\ln(4)/\ln(3)$. The \emph{image}
(under~$\bw$)
of the optimal subdivision found by the algorithm
is shown in figure~\ref{fig:opt-subdivision},
superimposed on the von Koch curve.
The evolution of the sum over the
\emph{normalized} number of iterations is shown in figure~\ref{fig:evol}.

The above description assumes that the value of $\alpha$ is the same as the
$\gamma$-dimension of the set $F$, say $\alpha_0$. We expect
$\delta$-independence in the values of $\sigma^\alpha[F, P(\delta)]$ where
$P(\delta)$ denotes the resultant subdivision of the algorithm at the scale
$\delta$, since the value of $\gamma^\alpha_\delta$ converges to a finite
nonzero value. This is what we observe from the values of $\sigma^\alpha[F,
P(\delta)]$ obtained for various values of $\delta$ (figure \ref{fig:evol}).

Now we would like to consider cases when $\alpha \neq \alpha_0$. Let
$0 < \delta_1 < \delta_2$. If $\alpha < \alpha_0$, then
$\gamma^\alpha(F, a, b) = \infty$. Therefore we expect that
$R(\alpha) =\sigma^\alpha[F, P(\delta_1)]$ $/\sigma^\alpha[F,P(\delta_2)] > 1$.
Similarly since $\alpha > \alpha_0$ implies $\gamma^\alpha(F,a,b) = 0$, we
expect that $R(\alpha) < 1$.

This fact can be used to algorithmically calculate the $\gamma$-dimension
$\alpha_0$: We need to find the number $\alpha_0$ such that $R(\alpha_0) =
1$. We already know that $\alpha_0 \in [1,m],\quad m$ being the embedding 
dimension,
 since $F \in \R^m$ is a curve. Treating
this as the initial bracket of values for $\alpha_0$, we just need to use
some algorithm such as bisection to shrink this bracket to sufficient
accuracy.
\section{The $F^\alpha$-Calculus} \label{sec:calculus}
Most of the proofs which are similar to the proofs in the case of discontinuous
sets like Cantor-like sets are omitted.
\subsection{$F$-Limit and $F$-Continuity}
\label{sec-f_continuity}
Now we introduce limits and continuity along a fractal curve.
\begin{definition} Let $F \subset \mathbf{R^n} $ be a fractal curve, and let
 $f:F \rightarrow \mathbf{R}$. Let $\theta \in F$. A number $l$ is said
to be the limit of $f$ throught points of $F$, or simply ${F}$\textbf{-
limit}, as 
$\theta' \rightarrow \theta$, if given $\epsilon > 0$ there exists 
$\delta > 0$ such that 
\[\theta' \in F \mbox{ and } |\theta' - \theta|<\delta 
\Longrightarrow |f(\theta') - l| 
< \epsilon\]

If such a number exists it is denoted by
\[\l = F\mbox{-}\lim_{\theta' \rightarrow \theta}f(\theta')\]
\end{definition}
\begin{definition}
A function $f : F \rightarrow R$ is said to be $\mathbf{F}$\textbf{-
continuous} at $\theta \in F$ if $f(\theta) = F \mbox{-}\lim_{\theta' \rightarrow \theta} f(\theta') $.
\end{definition}
\begin{definition}$f: F \rightarrow R$ is said to be uniformly continuous
on $E \subset F $ if for  any $\epsilon>0$ there exists $\delta>0$ such that
for any $\theta \in F$ and $\theta' \in E$ 
\[|\theta' - \theta|< \delta \Longrightarrow 
|f(\theta') - f(\theta)| < \epsilon \]
\end{definition}
\subsection{$F^\alpha$-Integration}
\label{sec-falpha-int}
We denote the class of bounded functions $f:F\rightarrow R$ by $B(F)$. 
\begin{definition}
For $t_1$,$t_2 \in [a_0,b_0]$,$t_1 \leq t_2$ a section or segment 
$C(t_1,t_2)$ of the 
curve  is defined as
\[ C(t_1,t_2) = \{\mathbf{w}(t') : t' \in [t_1,t_2] \} \]
\end{definition}
\begin{definition} 
Let $f:F \rightarrow \mathbf{R}$ and $t_1, t_2 \in [a_0,b_0]$,$t_1 \leq t_2$
and let
\[ M[f,C(t_1,t_2)] = \sup_{\theta \in C(t_1,t_2)} f(\theta)\]
and
\[ m[f,C(t_1,t_2)] = \inf_{\theta \in C(t_1,t_2)} f(\theta)\]

Let $S_F^\alpha(t)$ be finite for $ t \in [a,b] \subset [a_0
,b_0]$. Let $P =\{t_0, \ldots,t_n \}$ be a subdivision of $[a,b]$ . 
The upper and the lower $F^\alpha$-sum for the function $f$ over the
subdivision $P$ are given respectively by 
\be U^{\alpha} [f,F,P] =
\sum_{i=0}^{n-1}M[f,C(t_i,t_{i+1})][S_F^{\alpha}(t_{i+1})-S_F^{\alpha}(t_i)],
\label{eq:upper_sum}\ee
\be L^{\alpha} [f,F,P] =
\sum_{i=0}^{n-1}m[f,C(t_i,t_{i+1})][S_F^{\alpha}(t_{i+1})-S_F^{\alpha}(t_i)].
\label{eq:lower_sum}\ee
\end{definition}

Now we define the $F^\alpha$-integral
\begin{definition}
Let $F$ be such that $S_F^\alpha$ is finite on $[a,b]$. For $f \in B(F)$, 
the  lower and upper $F^\alpha$-integral of the function $f$ respectively 
, on the section $C(a,b)$ are
\be \underline{\int_{C(a,b)}} f(\theta) d_F^\alpha \theta =
\sup_{P_{[a,b]}}L^{\alpha}[f,F,P] \label{eq:upper_int} \ee
\be \overline{\int_{C(a,b)}} f(\theta) d_F^\alpha \theta =
\inf_{P_{[a,b]}}U^{\alpha}[f,F,P] \label{eq:lower_int}\ee

 If $f \in B(f)$, we say that $f$ is $F^\alpha$-
integrable on $C(a,b)$ if 
\[ \overline{\int_{C(a,b)}} f(\theta) d_F^\alpha 
\theta = \underline{\int_{C(a,b)}} f(\theta) d_F^\alpha \theta\]
and the common value is called the $F^\alpha$ -integral, denoted by 
\[\int_{C(a,b)} f(\theta) d_F^\alpha \theta. \]
\end{definition}
\subsection{$F^\alpha$-Differentiation}
\label{sec-alphadiff}
\begin{definition}\label{def:f-continue}
Let $F$ be a fractal curve. 
 Then the $F^\alpha$-derivative of function
 $f$ at $\theta \in F$ is defined as
\be (D_F^\alpha f)(\theta)= F \mbox{-}\lim_{\theta' \rightarrow \theta} 
\frac{f(\theta')-f(\theta)}
{J(\theta')-J(\theta)} \label{eq:derivative}\ee
if the limit exists.
\end{definition}
\begin{theorem}  If $(D_F^\alpha f)(\theta)$ exists for all $\theta
 \in C(a,b)$, then $f$ is $F$-continuous on $C(a,b)$.
\end{theorem}

\textit{Remark}: The $F^\alpha$-derivative $D^\alpha_F(f)$ of a constant 
function $f: F \rightarrow \mathbf{R}$, $f(\theta) = k \in R$ is zero.
This result is to be contrasted with the classical fractional derivative
(Riemann-Liouville, and others) of a constant, which is not zero in
general~\cite{Samko,Hilfer,Miller,Oldham}.

\subsection{Fundamental theorems of $F^\alpha$-calculus} \label{sec-fundamental}
The $F^\alpha$-integration and $F^\alpha$-differentiation are related as 
inverse operations
of each other. The first fundamental theorem states: 
\begin{theorem} \label{the:fundamental}
Let $f \in B(F)$ is an $F$-continuous function on $C(a,b)$,and let
$g:f\rightarrow \mathbf{R}$ be defined as
\[ g(\mathbf{w}(t)) = \int_{C(a,t)} f(\theta') d_F^\alpha \theta' \]
for all $t \in [a,b]$. Then
\[ (D_F^\alpha g)(\theta) = f(\theta)  \quad \mbox{ and } \theta = \bw(t) \]
\end{theorem}

The second fundamental theorem says that the $F^\alpha$-integral as a function
of upper limit is the inverse of $F^\alpha$-derivative except for an additive
constant.
\begin{theorem}Let $f:F\rightarrow \mathbf{R}$ be continuously $F^\alpha$-
differentiable function and $h :F \rightarrow \mathbf{R}$ be $F$-continuous, 
such that
$ h(\theta)  = (D_F^\alpha f)(\theta))$. Then
\[\int_{C(a,b)} h(\theta) d_F^\alpha \theta = f(\mathbf{w}(b)) - f(\mathbf{w}(a))\] \label{the:fundamental1} 
\end{theorem}
\subsection*{Comparison with approaches involving Local Fractional Operators}
The local fractional derivative (LFD) operator constructed in \cite{Chaos}
was based on the renormalization of
Riemann-Liouville differential operator on the real line. It was utilized to
establish the relation between the differentiability properties of nowhere
differentiable functions, such as Weierstrass function, and the dimension (Holder
exponent) of its graph. The domain of these functions is $\mathbf{R}$ and not 
a fractal. In \cite{Kolwankar} though 
the prescription was developed using LFD, it was realized that to make the local
fractional Fokker Planck equation causal and dynamically consistent, the
 evolution had
to be restricted to fractal subsets of the real line. Moreover the order of
differentiation had to be the dimension of the fractal support. Since Cantor-
like sets are totally disconnected, this necessitated a rigorous development of
 calculus on fractals from first principles, without using standard fractional
calculus on $\mathbf{R}$, which was carried out in \cite{Abhay}
The present paper is a logical extension of such a formulation, to fractal
curves. While Cantor-like sets considered in \cite{Abhay} are totally
 disconnected, the von-Koch like 
fractal curves are continuous but tangentless. Thus all the constructions and
proofs have to be carried out keeping in mind this difference of the domain of
 functions
and operators. (Thus for example the notion of 'set of change' and 
'$\alpha$-perfect sets' was crutial for cantor-like sets in \cite{Abhay} which
 is repleced by 
the invertibility of $S_F^\alpha$ in case of fractal curves considered in this
paper.)

There is multiplicity of approaches leading to the notion of local
fractional calculus. Various authors \cite{Jumarie}, \cite{Ben},\cite{Li} 
have further developed the notion
of local fractional differentiation with different approaches. They provide 
suitable framework for different classes of problems.
The development in \cite{Jumarie} is based on approach involving difference 
operators. In finding derivative
the quotient is taken with respect to $h^\alpha$ where $h$ is an increment of
the independent variable. 
This can be contrasted with the use of $S_F^\alpha$ in the present approach. 
This also reflects in the Taylor series where powers of
($S_F^\alpha$) appear (see equation (\ref{eq:tseries}) below) rather than powers of $h$ itself
as in \cite{Jumarie}.
 Further the domain is $\mathbf{R}$ in \cite{Jumarie} 
 whereas it is a fractal curve in the present paper.

In \cite{Ben} the notion of classical fractional
derivative is modified. Again the essential difference mentioned above for
Taylor series and the domain functions is also to be noted here.
The development in \cite{Li} is based on the Weyl Derivative and the 
domain of functions is $\mathbf{R}$ and not a fractal curve.

We may emphasize that in the present approach the role of the independent
 variables is 
delegated to the staircase/rise function $S_F^\alpha$, see e.g equation 
(\ref{eq:derivative}).
 In this sense our approach is like Stieltjes
approach in spirit but with some essential differences as noted in \cite{Abhay}. The function $S_F^\alpha$ captures the essence of 
fractal support, hence its use makes the calculus suitable for fractals.
\section{Conjugacy of $F^\alpha$-Calculus and Ordinary Calculus}
\label{sec:conjugacy}
In this section, we define a map $\phi$ which takes an $F^\alpha$-integrable 
function $f:F
\rightarrow \mathbf{R}$ to a Riemann integrable function $g:[S_F^\alpha(a_0),S_F^\alpha(b_0)]
\rightarrow \mathbf{R} $ such that their corresponding integrals have equal
values. Thus, the map $\phi$ exhibits a conjugacy between the two operations.

First let us define certain classes of functions:
\begin{enumerate}
\item $B(F)$ : class of bounded functions $f : F \rightarrow \mathbf{R}$.
\item $B([c,d])$ : class of bounded functions $f : [c,d] \rightarrow \mathbf{R}$

\item $\mathcal{L}(F)$: set of all functions which are $F^\alpha$-integrable on 
$C(a_0,b_0)$.

\item The image of $F$ under $S_F^\alpha$ is denoted by $K$, i.e $K = [S_F^\alpha(a_0),S_F^\alpha(b_0)]$, and $B(K)$ denotes the class of functions bounded on
 $K$.

\item $\mathcal{L}(K)$ denotes the class of functions in $B(K)$ which are 
Riemann integrable over the interval $ K = [S_F^\alpha(a_0),S_F^\alpha(b_0)]$.
\end{enumerate}

In order to fix the notation, here we briefly review the definition of 
Riemann integral. Firstly, if $g \in B([c,d])$ and $I \subset [c,d]$ is a 
closed interval, then we denote $M'[g,I] = \sup_{x \in I} g(x)$ and
$m'[g,I] = \inf_{x \in I} g(x)$. Further, the upper and lower sum over a 
subdivision $P_{[c,d]} = \{ y_0,\ldots,y_n \}$ are given by $U'[g,P] =
\sum_{i=0}^{n-1} M'[g,[y_i,y_{i+1}]]$ and $L'[g,P] = \sum_{i=0}^{n-1}
m'[g,[y_i,y_{i+1}]]$. If the upper and lower integrals given respectively
by $\inf_P U'[g,P]$ and $\sup_P L'[g,P]$ are equal, then $g$ is said to be 
Riemann integrable, and the Riemann integral
\[ \int_c^d g(y) dy\]
is defined to be the common value.

Now we define the above mentioned map $\phi$:
\begin{definition}The map $\phi : B(F)\rightarrow B([S_F^\alpha(a_0),
S_F^\alpha(b_0)])$ takes 
$f \in B(F)$ to $\phi[f] \in B([S_F^\alpha(a_0),S_F^\alpha(b_0)])$ such 
that for each $t \in [a_0,b_0]$,
\[\phi[f](S_F^\alpha(t))= f(\mathbf{w}(t))\]
\end{definition} \label{def:map}
\begin{lemma} \label{the:prophi}
The map $\phi: B(F) \rightarrow B(K)$ is one to one and onto.
\end{lemma}
The proof is straightforward. Thus we are assured 
that the inverse map $\phi^{-1}$ exists.

The following theorem brings out the conjugacy between $F^\alpha$-
integrals of functions along the fractal curve $F$ and the Riemann integrals
of their images under $\phi$.
\begin{theorem} A function $f\in B(F)$ is $F^\alpha$-integrable over
$C(a,b)$ if and only if $g= \phi[f]$ is Riemann integrable over
$[S_F^\alpha(a),S_F^\alpha(b)]$. In other words,a function $f \in B(F)$ 
belongs to $\mathcal{L}(F)$ if and only if $g \in \mathcal{L}(K)$. Further
\[\int_{C(a,b)} f(\theta) d_F^\alpha \theta = \int_{S_F^\alpha(a)}^{S_F^\alpha
(b)} 
g(u) du\] \label{the:foursix}
\end{theorem}
Proof: Let $ f: F \rightarrow R$ be $F^\alpha$-integrable. Then there exists a
subdivision $P_{[a,b]} = \{t_0,t_1,\ldots,t_n\}$ such that
\be U^\alpha [f,F,P] - L^\alpha [f,F,P] <\epsilon \label{eq:uminl}\ee  for any
$\epsilon >0$.

Denote $y_i = S_F^\alpha (t_i)$. Then $Q = \{y_i : 0 \leq i \leq n\}$ is a
subdivision of $[S_F^\alpha(a),S_F^\alpha (b)]$\\
For any component $[t_i,t_{i+1}]$
\begin{eqnarray*}
M[f,C(t_i,t_{i+1})] & = & \sup_{\mathbf{w} \in C(t_i,t_{i+1})} f(\mathbf{w})\\ 
		    & = & \sup_{t\in [t_i,t_{i+1}]} f(\mathbf{w}(t)) \\ 
		    & = & \sup_{t\in [t_i,t_{i+1}]} g(S_F^\alpha (t))\\
		    & = & \sup_{y\in [y_i,y_{i+1}]}g(y) \\
	            & = & M'[g,[y_i,y_{i+1}]] 
\end{eqnarray*}  

Therefore,
\begin{eqnarray}
U^\alpha[f,F,P]     & =&
\sum_{i=0}^{n-1}M[f,C(t_i,t_{i+1})][S_F^\alpha(t_{i+1})- S_F^\alpha(t_i)]
\nonumber\\
		    & =& \sum_{i=0}^{n-1}M[f,C(t_i,t_{i+1})][y_{i+1}-y_i]
\nonumber\\
		    & =& \sum_{i=0}^{n-1}M'[g,[y_i,y_{i+1}]][y_{i+1}-y_i]
\nonumber\\
		    & =& U'[g,Q] \label{eq:U'_gQ} 
\end{eqnarray} 
Similarly
\be L^\alpha[f,F,P]= L'[g,Q] \label{eq:L'_gQ}\ee
then using equations (\ref{eq:uminl}), (\ref{eq:U'_gQ}) and  (\ref{eq:L'_gQ})
\[ U'[g,Q]- L'[g,Q]< \epsilon \]
which implies that $g$ is Riemann integrable over
$[S_F^\alpha(a),S_F^\alpha(b)]$ and \\
\[\int_{S_F^\alpha(a)}^{S_F^\alpha(b)} g(u) du = \int_{\theta} f(\theta) 
d_F^\alpha \theta
\]
Conversely if $g$ is Riemann Integrable, then for given $\epsilon>0$ there
exists a subdivision $Q' = \{v_0,\ldots,v_m\}$ of
$[S_F^\alpha(a),S_F^\alpha(b)]$ such that $U'[g,Q']-L'[g,Q'] <\epsilon$.
Then the converse can be proved by following the above steps in the reverse
order.

$\bullet$

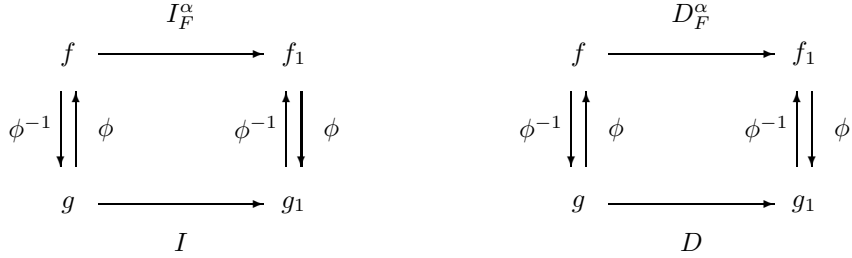
\begin{figure}[htb]
\noindent
\begin{minipage}{0.4\textwidth}%
\setlength{\unitlength}{1cm}
\newcommand{\cf}[1]{\makebox(0,0){$#1$}} 
\begin{picture}(6,4)(0,0)
\put(1,1){\cf{g}}
\put(1,3){\cf{f}}
\put(4,1){\cf{g_1}}
\put(4,3){\cf{f_1}}
\put(1.5,2){\cf{\phi}}
\put(0.5,2){\cf{\phi^{-1}}}
\put(2.5,0.5){\cf{I}}
\put(2.5,3.5){\cf{I^\alpha_F}}
\put(3.5,2){\cf{\phi^{-1}}}
\put(4.5,2){\cf{\phi}}
\put(0.9,2.5){\vector(0,-1){1}}
\put(1.1,1.5){\vector(0,1){1}}
\put(1.4,1){\vector(1,0){2.2}}
\put(3.9,1.5){\vector(0,1){1}}
\put(4.1,2.5){\vector(0,-1){1}}
\put(1.4,3){\vector(1,0){2.2}}
\end{picture}%
\end{minipage}%
\hspace{0.15\textwidth}
\begin{minipage}{0.4\textwidth}%
\setlength{\unitlength}{1cm}
\newcommand{\cf}[1]{\makebox(0,0){$#1$}} 
\begin{picture}(6,4)(0,0)
\put(1,1){\cf{g}}
\put(1,3){\cf{f}}
\put(4,1){\cf{g_1}}
\put(4,3){\cf{f_1}}
\put(1.5,2){\cf{\phi}}
\put(0.5,2){\cf{\phi^{-1}}}
\put(2.5,0.5){\cf{D}}
\put(2.5,3.5){\cf{D^\alpha_F}}
\put(3.5,2){\cf{\phi^{-1}}}
\put(4.5,2){\cf{\phi}}
\put(0.9,2.5){\vector(0,-1){1}}
\put(1.1,1.5){\vector(0,1){1}}
\put(1.4,1){\vector(1,0){2.2}}
\put(3.9,1.5){\vector(0,1){1}}
\put(4.1,2.5){\vector(0,-1){1}}
\put(1.4,3){\vector(1,0){2.2}}
\end{picture}%
\end{minipage}%
\caption{The relation between $F^\alpha$-integral and Riemann integral,
also between $F^\alpha$-derivative and Ordinary derivative}%
\label{fig:map-c}%
\end{figure}

Let $f_1$ denote the indefinite $F^\alpha$-integral viz.
$ f_1(\mathbf{w}(t)) = \int_{C(a,t)} f(\theta) d_F^\alpha \theta $
and let $g_1$ denote the ordinary indefinite Riemann integral viz.
$ g_1(y) = \int_{S_F^\alpha(a)}^{y} g(y') dy' $.
If we further denote the indefinite $F^\alpha$-integral operator  by $I_F^\alpha$ and the indefinite Riemann integral operator by $I$, then the result of 
theorem (\ref{the:foursix}) can be expressed as 
\[ I_F^\alpha = \phi^{-1} I \phi \]
 as displayed in the commutative diagram of figure \ref{fig:map-c}.

The following theorem brings out the conjugacy between $F^\alpha$-derivative
and ordinary derivative.
\begin{theorem}
Let $h$ be a function in $B(F)$ such that $g= \phi [h]$ is ordinarily 
differentiable on $K =$ range of $S_F^\alpha$. Then $D_F^\alpha h(\theta)$ 
exists for all $\theta \in F $ and
\[D_F^\alpha h(\theta) = \frac{dg(v)}{dv}|_{v= J(\theta)}\]
\end{theorem}
Proof: Let $v \in K$. Then by definition
\[\frac{dg}{dv} = \lim_{u\rightarrow v}\frac{ g(u) - g(v)}{u - v}\]
i.e given $\epsilon_0 > 0$, there exists $\delta_0 > 0$ such that
\[|u-v| < \delta_0 \Longrightarrow |\frac{dg}{dv} - \frac{g(u) - g(v)}{u-v}|
< \epsilon_0 \]
Let us recall our assumption that $S_F^\alpha$ is monotonically increasing and 
one-to-one.
Let $t= (S_F^\alpha)^{-1} (v)$, $t' = (S_F^\alpha)^{-1} (u)$. Then $t,t' 
\in [a_0,b_0]$,$h(\mathbf{w}(t')) = g(u)$
and $h(\mathbf{w}(t)) = g(v)$. Thus,
\[|S_F^\alpha(t') - S_F^\alpha(t)| < \delta_0 \Longrightarrow |\frac{dg}{dv}
- \frac{h(\mathbf{w}(t')) - h(\mathbf{w}(t))}{S_F^\alpha(t') - S_F^\alpha(t)}|
< \epsilon_0 \]
Since $(\mathbf{w})^{-1}$ and $S_F^\alpha$ are continuous, so is their 
composition $S_F^\alpha \circ (\mathbf{w})^{-1}$. Therefore, there exists 
$\delta_1 > 0$ 
such that  
\[|\mathbf{w}(t') - \mathbf{w}(t)| < \delta_1 \Longrightarrow |S_F^\alpha(t')
- S_F^\alpha(t)| < \delta_0 \]
\[\Longrightarrow |\frac{dg}{dv} - \frac{h(\mathbf{w}(t') - h(\mathbf{w}(t))}
{S_F^\alpha(t') - S_F^\alpha(t)}| < \epsilon_0. \]
Setting $\theta' = \bw(t'), \theta = \bw(t)$, we can rewrite this as
\[|\theta' - \theta | < \delta_1 \Rightarrow |J(\theta') - J(\theta)| < 
\delta_0 \Rightarrow |\frac{dg}{dv} - \frac{h(\theta') - h(\theta)}{J(\theta')
- J(\theta)}| < \epsilon_0\]
which by definition of $F$-limit and $D_F^\alpha$ means
\[D_F^\alpha h(\theta) = \Flim_{\theta' \rightarrow \theta
} \frac{h(\theta') - h(\theta)}{J(\theta') - J(\theta)} =
\frac{dg}{dv}|_{v=J(\theta)}.\]
$\bullet$
\begin{theorem}
Let $h \in B(F)$ be an $F^\alpha$-differentiable function at all $\theta \in
F$. Further, let $g= \phi[h]$. Then $dg/dv$ exists at $v= J(\theta)$ and
\[\frac{dg(v)}{dv}|_{v=J(\theta)} = D_F^\alpha h(\theta)\]
\end{theorem}
Proof: As $g=\phi[h]$, we have $g(S_F^\alpha(t)) = h(\mathbf{w}(t))$ for all 
$t\in [a_0,b_0]$  i.e  $g(J(\theta)) = h(\theta)$ for all $\theta \in F$.

By definition and substitution
\begin{eqnarray*}
D_F^\alpha h(\theta) &= & \Flim_{\theta' \rightarrow \theta}
\frac{h(\theta') - h(\theta)}{J(\theta') - J(\theta)}\\
&=& \Flim_{\theta' \rightarrow \theta} \frac{g(J(\theta')) 
-g(J(\theta))}{J(\theta') - J(\theta)}
\end{eqnarray*}
Thus given $\epsilon_0> 0$ there exists ${\delta_0}' > 0$ such that
\[|\theta' - \theta| < {\delta_0}' \Longrightarrow 
| \frac{g(J(\theta')) -g(J(\theta))}{J(\theta') - J(\theta)}
-D_F^\alpha h(\theta)| < \epsilon_0.\]
Let $v = J(\theta)$ and $u=J(\theta')$, i.e $\theta = J^{-1}(v) $ and $\theta'
= J^{-1}(u)$. Since $J^{-1}$ is continuous, there exists a $\delta > 0$ such 
that
\[|u-v| < \delta \Longrightarrow | \theta' - \theta| < \delta_0 
\Longrightarrow |\frac{g(u)-g(v)}{u-v}-D_F^\alpha h(\theta)| < \epsilon_0 \]

Which by definition of ordinary derivative gives
\[\frac{dg}{dv}|_{v=J(\theta)} = \lim_{u\rightarrow v} \frac{g(u) - g(v)}
{u-v} = D_F^\alpha h(\theta)\]
$\bullet$

This conjugacy can also be expressed as $D_F^\alpha = \phi^{-1} D \phi$ 
as shown in the commutative diagram of figure \ref{fig:map-c}.

\subsection{Taylor Series}
 One can write a fractal Taylor series for functions on fractal
 curve $F$, by
using the results of this section. 

If $g= \phi[h]$ be such that the ordinary Taylor series is given by
\[g(u) = \sum_{n=0}^\infty \frac{(u -y)^n}{n!} \frac{d^n g(y)}{dy^n}\]
is valid for $u,y \in [S_F^\alpha(a),S_F^\alpha(b)]$, then for $\theta,\theta'
\in F$ it can be seen that 
\be \label{eq:tseries}
h(\theta) = \sum_{n=0}^{\infty} \frac{(J(\theta)-J(\theta'))^n}{n!} 
(D_F^\alpha)^n h(\theta')
\ee
provided $h \in B(F)$ is $F^\alpha$- differentiable any number of times on
 $C(a,b)$
such that $(D_F^\alpha)^n h \in B(F)$ for any integer $n > 0$.

\subsection{Function Spaces in $F^\alpha$-Calculus}\label{sec:funcspace}
We introduce the following spaces:

The set of all functions that have $F$-continuous $F^\alpha$-derivatives upto
order $k$ can be analysed analogous to \cite{Abhay00} and are defined by
$C^k(F), k \in \mathbf{N}$:Set of all functions $f:F \rightarrow 
\mathbf{R}$ such that

\[(D_F^\alpha)^n f \in C^0(F) \mbox{ for all } n \leq k\]
One can define norm on $C^k(F)$ for $F \subset \R^n$, similar to what is defined for $F\subset \R$ using the $F^\alpha$-derivative as follows:
\[||f|| = \sum_{0 \leq n \leq k} \sup_{\theta \in F} |[(D_F^\alpha)^n f]\theta
| \quad f \in C^k(F)\]
$C^k(F)$ are complete with repect to this norm. Unlike in \cite{Abhay00}the
 need
for $S_F^\alpha$-concordant functions does not arise since $S_F^\alpha(u) \neq
S_F^\alpha(v)$ even if $f(\bw(u)) = f(\bw(v))$ for $u \neq v$. It can also be 
shown that quite easily that $C^k(F)$ is separable.

The spaces of $F^\alpha$-integrable functions and their completion can also be
constructed in an analogous manner as is done in \cite{Abhay} for fractal 
subsets of real line.

Set $\mathcal{L}(F)$ of $F^\alpha$-integrable functions is a vector space with
usual operations of addition and scalar multiplication. An appropriate norm 
$\mathcal{N}_p$ can be defined for $F^\alpha$ integrable functions which
 satisfies all the required properties.
\[\mathcal{N}_p (f) = ||f||_p = [\int_{C(a,b)} |f(\theta)|^p d_F^\alpha \theta]^{1/p} \quad 1 \leq p < \infty \] 
$\mathcal{N}_p $ can be shown to act as a norm on $L'(F)$ where $L'(F)$ is a
 vector space of equivalent classes of $\mathcal{L}(F)$, the class of all
 $F^\alpha$-
integrable functions. $L_p'(F)$ which is $L'(F)$ with specific norm $\mathcal{N
}_p$ is not complete but can be completed using standard procedure. The complete
space is then denoted by $L_p(F)$ and is a Banach space. The spaces $L_p'(F)$
and $L_p(F)$ can also be shown to be separable.

Analogues of abstract Sobolev Spaces can be constructed in exactly the same
 way as is done in the above cited reference for subsets of Real line.

\section{Example: Absorption on fractal curves }
\label{sec:absorption}
Consider the flux of a fluid or  of particles  moving steadily through and
 getting
 absorbed in a percolating cluster or fractured rock. A simple mathematical 
model of this process for a single branch would be that of particles getting
 absorbed along a 
fractal path. The absorption process can then  be modelled by the following 
equation:
\begin{equation} \label{eq:absorption}
D_F^\alpha \rho(\theta) = - \kappa \rho(\theta)
\end{equation} 
where $\rho(\theta)$ is the density of fluid at a point $\theta$ of the fractal
 channel (e.g. backbone of the percolating cluster), $\kappa$ being the 
coefficient of absorption (which in a simple model is taken as constant),  
and  $D_F^\alpha$ is
 the $F^\alpha$ - derivative.

The left hand side of equation (\ref{eq:absorption})represents the space rate
 of 
change of density of particles at a particular position on the fractal curve
(or path) varying along the path. 

The exact solution of the above equation can be obtained using the conjugacy
 between $F^\alpha$-derivative and ordinary derivative and applying the corresponding operator on $\rho(\theta)$  as follows:
\[\phi \rho(\theta) = \tilde{\rho}(y) \mbox{ where } y = J(\theta)\]
then equation (\ref{eq:absorption}) becomes
\[\frac{d}{dy} \tilde{\rho}(y) = - \kappa \tilde{\rho} (y) \]
the solution of which is given by
\[\tilde{\rho}(y) = \tilde{\rho}(0) \exp(- \kappa y)\]
Applying the inverse conjugate operator to the above equation we obtain
\[
\rho(\theta) = \rho(0) \exp(-\kappa J(\theta))
\]
or since $\theta = \bw(u)$ and $J(\theta) = S_F^\alpha(u)$,
\be
\rho(\mathbf{w}(u)) = \rho(0) \exp(-\kappa S_F^\alpha(u))
\ee 
This is like stretched exponential behaviour in view of the relation between 
euclidean distance and staircase (see fig ~\ref{fig:graph2}).
\section{Conclusion}
\label{sec:conclusion}
In this paper we have developed a calculus on parametrizable fractal curves of 
dimension $\alpha \in [1,n]$. This involves the identification of the 
important role played by the mass function and the corresponding (rise)
staircase function which may be 
compared with the role played by the independent variable itself in ordinary 
calculus. The definitions of $F^\alpha$-integral and $F^\alpha$-derivative are 
specifically tailored for fractal curves of dimension $\alpha$. Further they
 reduce to Riemann integral and ordinary derivative respectively, when $F = 
\mathbf{R}$ and $\alpha =1$.

Much of the development of this calculus is carried in analogy with the 
ordinary calculus. Specifically, we have adopted Riemann-Stieltjes like
approach for integration, as it is direct, simple and advantageous from 
algorithmic point of view. 
The example of absorption on fractal curves mentioned in section 
\ref{sec:absorption} demonstrates the utility of such a framework in modelling.
Other applications may include fractal Langevin
equation for Brownian motion and Levy processes on such curves, which will 
follow in future work.
This approach may be further
useful in dealing with path integrals and other similar applications.
Another direction for extension of the considerations in this paper is the 
 extension to  crumpled
or fractal surfaces which are continuously parametrizable by a finite number
 of
 variables.
\section*{Acknowledgements}
Seema Satin is thankful to Council of scientific and Industrial Research (CSIR) India
for financial assistance.

\end{document}